\documentclass[aps,physrev,reprint,amsmath,amssymb]{revtex4-2}

\usepackage{siunitx}
\usepackage{graphicx}%
\usepackage{amsmath,amssymb,amsfonts}%
\usepackage{amsthm}%
\usepackage{mathrsfs}%
\usepackage{booktabs}%
\usepackage{hyperref}
\usepackage{listings}%
\usepackage{multirow}
\usepackage[toc,page]{appendix}

\newcommand{\tauon}{\tau_\mathrm{on}}
\newcommand{\tauth}{\tau_\mathrm{thermal}}
\newcommand{\Tc}{T_\mathrm{C}}
\newcommand{\Tsub}{T_\mathrm{SUB}}
\newcommand{\TCh}{T_\mathrm{Ch}}
\newcommand{\wc}{w_\mathrm{c}}
\newcommand{\wh}{w_\mathrm{h}}
\newcommand{\Rsheet}{R_\mathrm{SHEET}}

\newcommand{\Isw}{I_\mathrm{SW}}
\newcommand{\Jsw}{J_\mathrm{SW}}
\newcommand{\Jswtilde}{\widetilde{J}_\mathrm{SW}}
\newcommand{\Jideal}{\widehat{J}_\mathrm{SW}}
\newcommand{\Jmodel}{J_\mathrm{MODEL}}
\newcommand{\Jconstr}{J_\mathrm{CONSTR}}
\newcommand{\Jc}{\widehat{J}_\mathrm{C}}
\newcommand{\Ih}{I_\mathrm{H}}
\newcommand{\Ihtilde}{\widetilde{I}_\mathrm{H}}
\newcommand{\ih}{i_\mathrm{H}}
\newcommand{\Ir}{I_\mathrm{R}}
\newcommand{\Isupp}{I_{\mathrm{H,SUPP}}}

\usepackage{setspace}

\begin{document}


\title{Parameter extraction for a superconducting thermal switch (hTron) SPICE model}


\author{Valentin Karam}
\email[Corresponding author: ]{vkaram@mit.edu}
\author{Owen Medeiros}
\author{Tareq El Dandachi}
\author{Matteo Castellani}
\author{Reed Foster}
\author{Marco Colangelo}
\author{Karl Berggren}

\affiliation{Department of Electrical Engineering and Computer Science, Massachusetts Institute of Technology, 77 Massachusetts Ave., Cambridge, 02139, Massachusetts, USA.}


\date{August 12, 2024}

\begin{abstract}
Efficiently simulating large circuits is crucial to the development of superconducting nanowire-based electronics. However, current simulation tools for this technology are not adapted to the scaling of circuit size and complexity.
We focus on the multilayered heater-nanocryotron (hTron), a promising superconducting nanowire-based switch used in applications such as superconducting nanowire single-photon detector (SNSPD) readout. Previously, the hTron was modeled using traditional finite-element methods (FEM), which fall short in simulating systems at a larger scale. An empirical-based method would be better adapted to this task, enhancing both simulation speed and agreement with experimental data. In this work, we perform switching current and activation delay measurements on 17 hTron devices. We then develop a method for extracting physical fitting parameters used to characterize the devices. We build a SPICE behavioral model that reproduces the static and transient device behavior using these parameters, and validate it by comparing its performance to a model developed in prior work, showing an improvement in simulation time by several orders of magnitude. Our model provides circuit designers with a tool to help understand the hTron’s behavior during all design stages, thus enabling broader use of the hTron across various new areas of application.

\end{abstract}


\maketitle

\section{Introduction}
\label{sec:Section1}

The field of superconducting electronics has demonstrated significant benefits in areas such as quantum computing \cite{ladd2010quantum}, imaging \cite{oripov2023superconducting}, neuromorphic computing \cite{islam2023review}, and digital logic \cite{Likharev1993}. Josephson junction (JJ)-based circuits dominate the field, thanks to their fast operation speeds and low power dissipation \cite{soloviev2017beyond}. However, JJs have low signals, low normal resistance, and are sensitive to magnetic fields. In contrast, cryotrons based on superconducting nanowires can operate in noisy environments, offer high gains, and exhibit high impedance and fan-out \cite{mccaughan2014superconducting, zheng2020multigate, huang2023splitter}. Therefore, cryotrons present a complementary device to JJs for certain applications. In particular, the capability of superconducting nanowire-based circuits to drive high loads makes them compatible with complementary metal-oxide-semiconductor (CMOS) technologies \cite{zhao2017nanocryotron, toomey2019bridging, mccaughan2019superconducting}.

Multiple superconducting nanowire-based devices have been introduced in the past, such as the nanocryotron (nTron) \cite{mccaughan2014superconducting} and the heater-nanocryotron (hTron) \cite{baghdadi2020multilayered}. The nTron is a three-terminal device that uses a constriction at the gate input to suppress a superconducting channel. It has a maximum demonstrated clock frequency of $\SI{615.4}{\MHz}$ \cite{zheng2019characterize}, but suffers from leakage current between the gate and channel. The hTron device --- in its multilayered version --- uses the Joule heating from a normal heater to suppress a superconducting channel deposited beneath it. An oxide layer separates the heater layer from the superconducting layer, which isolates the layers galvanically while coupling them thermally. The hTron is in some ways easier to fabricate than the nTron due to the absence of a narrow constriction (although it does benefit from the availbility of multiple layers), and does not exhibit measurable leakage currents \cite{baghdadi2020multilayered}.

These two nanowire-based devices enabled the creation of a logic family \cite{buzzi2023nanocryotron}, a memory cell \cite{butters2021scalable}, and are particularly used in superconducting nanowire single-photon detector (SNSPD) arrays, both for pulse amplification and readout \cite{ zheng2020superconducting, foster2023superconducting, castellani2024nanocryotron}. The nTron can also be used to translate SFQ pulses to CMOS \cite{zhao2017nanocryotron}. When a higher impedance is needed, an nTron and hTron can be combined in an amplification stage, \textit{e.g.}, to drive an LED and communicate between superconducting neurons \cite{shainline2019superconducting, castellani2020thesis}. If used in larger and more complex circuits,  superconducting nanowires thus have the potential to complement Josephson junctions (JJs) and thus enable new superconducting electronics applications.

However, the scaling of these nanowire-based circuits is currently limited \cite{castellani2024nanocryotron, foster2023superconducting, huang2024monolithic}. We can partly explain this deficiency by the lack of simulation tools for superconducting nanowires, which is crucial to their development. Baghdadi et al. developed a 3D electrothermal model for the hTron using finite-element modeling (FEM) techniques, modeling heat exchanges inside the device by solving differential heat equations \cite{baghdadi2020multilayered}. This model, while helpful at the device level during early development stages, cannot easily simulate large-scale circuits, mainly because of the stiffness of the differential equations and the difference of scales in the geometry of these layered devices. A simpler, less accurate 0D model has also been developed, which was implemented in SPICE by Castellani \cite{castellani2020thesis}. However, this simpler model is also slow to solve and suffers from convergence issues. Moreover, both models exhibit imperfect agreement with experimental data. Indeed, the heat equations require various physical parameters to be approximated from literature or experiment \cite{baghdadi2020multilayered}. This means that the model has to be tuned and optimized each time the geometry is changed. An arbitrary heater and channel width cannot simply be plugged in to get a close alignment with measurements, because many of the material's thermal parameters are geometry-dependent (\textit{e.g.}, boundary resistance or diffusion coefficient).

These issues can be solved by developing a physics-informed behavioral model. Instead of focusing on the microscopic physics and heat exchanges within the device, behavioral models fit experimental data using a minimal set of parameters. This approach allows physics-informed behavioral models to be simpler due to their empirical basis, while also being robust thanks to fitting equations arising from phenomenological explanations of the underlying physics.

More precisely, we can divide the hTron response to a given heater-current pulse into a static and transient response. The static response is defined by the channel switching current measured at DC, whereas the transient response is defined by the device activation delay --- delay between the input of a heater pulse and the channel switching from superconductive to resistive. To model the hTron behavior, the critical current and activation delay dependence on heater current thus have to be measured.

In this paper, we characterized 17 hTron devices from a single wafer and developed a model for their static behavior that requires only two physical parameters. We systematically extracted these parameters from measurements of critical current as a function of heater current. Furthermore, we correlated these parameters with the heater and channel widths, allowing us to estimate them from the device geometry. The transient response of the device, which depends on the heat flow from the heater to the channel through the oxide layer, was also modeled to fit experiments, allowing us to accurately simulate the device close to the maximum operating speed. Finally, we implemented an hTron behavioral model in LTSpice and assessed its simulation speed and accuracy by comparing it to the previous electrothermal model. This was conducted by applying our parameter extraction method to measurement data from the earlier hTron study \cite{baghdadi2020multilayered}. We achieved improved agreements with published experimental data and successfully replicated published results. Moreover, our simulation time was faster by several orders of magnitude, making our approach compatible for use by circuit designers.

\section{Methods}
\label{sec:Section2}

\begin{figure*}[t]
    \includegraphics[width=\textwidth]{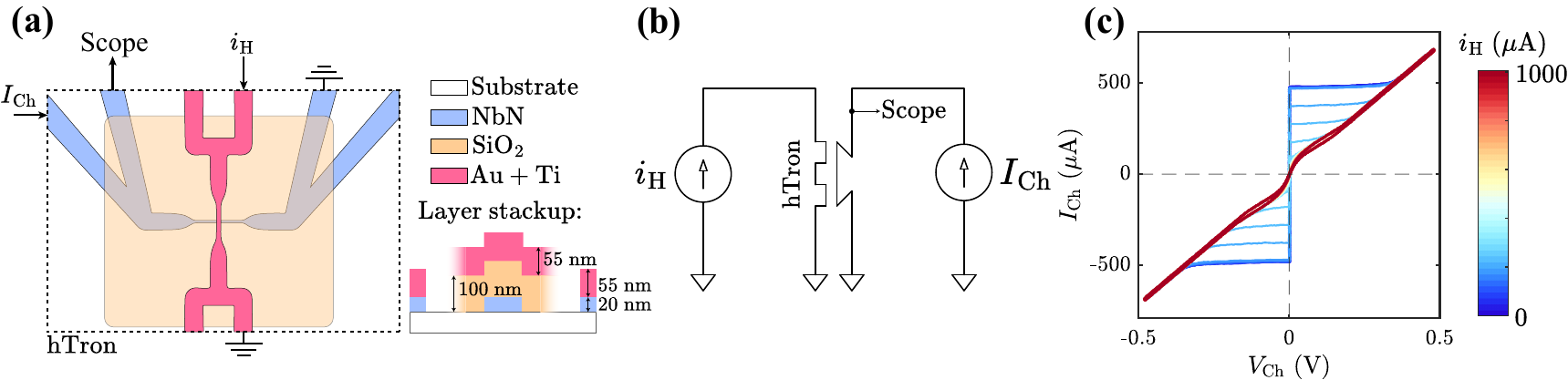}
    \caption{\textbf{Geometry, typical circuit and I-V curves of hTron device.} \textbf{(a)} Top-down schematic view of device geometry showing the various material layers, as well as cross-section showing the stack-up. The channel width (in blue) and the heater width (in red) are $\SI{100}{\nm}$ and $\SI{500}{\nm}$, respectively. Labels indicate terminals for electrical contacts; \textbf{(b)} electrical schematic diagram showing hTron bias and measurement scheme; and \textbf{(c)} I-V curves of channel current against channel voltage for various heater current $\ih$. The switching current is reduced by the application of a heater current.}
    \label{fig:methods_geometry}
\end{figure*}

This section outlines our methodological approach to constructing, characterizing, and modeling hTron devices.

\subsection{Fabrication}

Here we present the hTron fabrication process we used, which is comparable to the one introduced by Baghdadi et al \cite{baghdadi2020multilayered}. Figure \ref{fig:methods_geometry}a depicts the device layout together with the layer stack-up. 

The hTron fabrication process started with the deposition of a uniform layer of Niobium nitride (NbN) layer with thickness $d_\text{c}=\SI{20}{\nm}$ onto a silicon substrate using RF-biased sputtering \cite{dane2017bias}. Subsequently, $\SI{5}{\nm}$ Ti and $\SI{50}{\nm}$ Au were deposited with a liftoff process to pattern the pads, that served as wire-bonding connections for the hTron ports. Following this, the channel was patterned into the underlying NbN layer with electron beam lithography (EBL) followed by reactive ion etching. A $\SI{100}{\nm}$ silicon oxide layer was deposited using a PECVD technique to isolate the heater and channel layers. Finally, the heaters were patterned with another liftoff process using the same Ti/Au thickness ratio.

In total, 9 distinct geometry combinations were patterned on two $10\times\SI{10}{\mm}$ dies from a single wafer. The nanowire configurations varied in heater width $\wh$ and channel width $\wc$ of $\SI{100}{\nm}$, $\SI{500}{\nm}$, and $\SI{1}{\um}$, and all had lengths of $\SI{10}{\um}$. Though two identical dies (18 devices) were measured, one device with $\wh=\SI{100}{\nm}$ and $\wc=\SI{1}{\um}$ was damaged while measuring, resulting in a total of 17 functional devices. As seen in Figure \ref{fig:methods_geometry}a, each terminal of the hTron was connected to 2 different pads, enabling a 4-point measurement setup.

\subsection{Measurement Setup}

This section details the experimental setup used to measure switching current $\Isw$ as a function of heater current $\ih$, $\Isw(\ih)$, as shown in Figure \ref{fig:methods_geometry}a and Figure \ref{fig:methods_geometry}b. This relation explains the steady-state behavior of the hTron device. For our experiments, we used an immersion probe inside a liquid helium dewar \cite{butters2022digital}, setting a substrate temperature of $\Tsub=\SI{4.2}{\K}$.

The switching current was determined by gradually sweeping the channel current both positively and negatively using an arbitrary waveform generator (AWG) until the device switched into the resistive state, resulting in a detectable voltage. The AWG waveform frequency was set to $\SI{10}{\kHz}$. Throughout this paper, we will use the switching current density $\Jsw=\Isw/(\wc\cdot d_\text{c})$ instead of the switching current $\Isw$, making our measurements independent of the channel width. The dependence of the switching current on the heater current was then obtained by biasing the heater with a DC current $\ih$ while performing the channel sweep, locally increasing the channel temperature. A typical result is shown as a current-voltage (I-V) plot in Figure \ref{fig:methods_geometry}c, where we can see the switching current being gradually lowered by the increase of the heater current. For high heater currents, the channel temperature approached the film critical temperature $\Tc$, and thus the channel was fully suppressed. To account for the intrinsic stochastic nature of the switching characteristics of the devices, we averaged 100 trials for each datapoint \cite{lin2013role}.

The channel switching current can also be modulated by increasing the entire substrate temperature. To characterize this effect, we used a commercial temperature controller with a heater located inside the immersion probe. By doing so, the whole sample was globally heated, allowing the characterization of the direct dependence between the channel's switching current and temperature.

\subsection{Modeling}

This subsection presents our approach to modeling the hTron device, focusing on developing a behavioral model grounded in empirical data. We describe the fitting functions and parameters, critical for accurately replicating and predicting the hTron device's behavior.

\subsubsection{Physics-informed behavioral model}
We developed a physics-informed behavioral model based on a set of equations and fitting parameters to fit the $\Isw(\ih)$ experimental data that we collected. In our approach, we treated the hTron as a four-terminal black box, focusing on modeling its behavior rather than its intricate physical properties, which were unknown.

The fitting parameters varied from one device to another, and across two devices with the same geometry. However, we found correlations between the fitting parameters and the device widths ($\wc$ and $\wh$), enabling prediction of the behavior of a device from its geometry.

\subsubsection{Fitting functions}

We modeled the hTron static behavior with two analytic expressions: the first estimates the channel temperature from a heater current input, and the second predicts the switching current at this temperature.
To estimate the channel temperature $\TCh$ from the heater current, we used the expression 

\begin{equation} \label{eq:T_vs_Ih}
\TCh(\ih) = \left[(\Tc^4 - \Tsub^4)\cdot  \left(\frac{\ih}{\Isupp}\right)^\eta + \Tsub^4\right]^{\frac{1}{4}}
\end{equation}

where $\Isupp$ is the suppressing current --- the heater current at which the channel is fully suppressed and reaches the critical temperature $\Tc$; and $\eta$ is the strength of the thermal dependence between the heater current and the estimated channel temperature. We used the value of $\eta=2$ in this paper based on its optimal fit with Baghdadi et al.'s measurement data \cite{baghdadi2020multilayered}. However, Butters suggests $\eta=3$ (See Equation 3.30 of \cite{butters2022digital}), implying a stronger dependence of the channel temperature with the heater current. More details about our choice of $\eta$ can be found in the Supplemental Material.

We then found the unconstricted switching current density from the estimated channel temperature, $\Jideal(\TCh)$, defined as the switching current density of the channel section located immediately below the heater:

\begin{equation}\label{eq:J_vs_Tnw}
\begin{aligned}
& \Jideal(\TCh)=\Jc \cdot \left[1-\left(\frac{\TCh}{\Tc}\right)^3\right]^{2.1} \mathrm{,} \quad \TCh\leq\Tc\\
& \text{with } \Jc=\frac{\Jideal(\ih=0)}{\left[1-\left(\frac{\Tsub}{\Tc}\right)^3\right]^{2.1}} 
\end{aligned}
\end{equation}

This function was introduced in \cite{baghdadi2020multilayered}. The parameter $\Jc$ is the channel's unconstricted critical current density, \textit{i.e.}, the switching current at zero Kelvin of the channel section located immediately below the heater. It can be obtained directly from $\Jideal(\ih=0)$, the unconstricted switching current density at zero heater current (\textit{i.e.}, at substrate temperature $\Tsub$).

By combining these two simple functions in the form $\Jideal(\TCh(\ih))= \Jideal(\ih)$, we were able to accurately fit our measurement data $\Jsw(\ih)$. The fitting parameters $\Jideal(\ih=0)$ and $\Isupp$ were extracted from measurements. 

\begin{widetext}
\begin{equation}
\label{eq:Jideal}
\begin{aligned}
\Jideal(\ih) = \Jc \cdot \left[1-\left(\frac{1}{\Tc}\cdot\left[(\Tc^4 - \Tsub^4)\cdot  \left(\frac{\ih}{\Isupp}\right)^\eta + \Tsub^4\right]^{\frac{1}{4}}\right)^3\right]^{2.1}
\end{aligned}
\end{equation}
\end{widetext}

Equation \ref{eq:Jideal} simply combines Equation \ref{eq:T_vs_Ih} and Equation \ref{eq:J_vs_Tnw}. It models the switching current density of the channel portion located directly below the heater. However, at low heater currents, the devices often exhibited a constant switching current, or plateau. This plateau suggests the presence of a defect or constriction along the channel in an area distant from the heater (as detailed in Section \ref{sec:Section3}). In that case, the measured switching current density at substrate temperature would be smaller than $\Jideal(\ih=0)$. To match these observations, the model switching current density can be rewritten as:

\begin{equation}\label{eq:Jmodel}
\begin{aligned}
& \Jmodel(\ih)=\mathrm{min}\left\{\Jideal(\ih),\Jconstr\right\}, \\ & \Jconstr\leq\Jideal(\ih=0),
\end{aligned}
\end{equation}

where the value $\Jconstr$ is the measured switching current when the heater current is zero, \textit{i.e.}, at substrate temperature. A value of $\Jconstr=\Jideal(\ih=0)$ would mean no plateau was observed. Most of our measured devices presented a plateau at low heater currents, which could be as low as half the ideal switching current at substrate temperature $\Jideal(\ih=0)$.

\begin{figure*}[t]
    \centering
    \includegraphics[width=\textwidth]{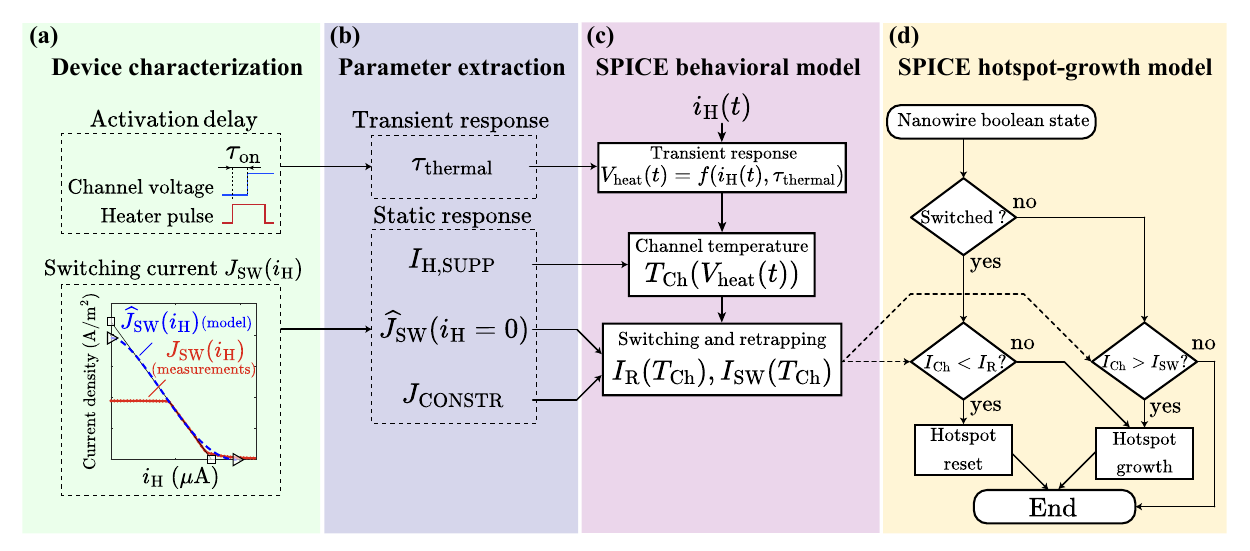}
    \caption{\textbf{Diagram of end-to-end process for the hTron SPICE modeling.} \textbf{(a)} Device characterization consisting of activation delay measurements and switching current against heater current measurements. Activation delay measurements provide the transient device response, and heater current-dependent switching current measurements define the static device response. \textbf{(b)} Extraction of key model parameters: the heat transfer time constant $\tauth$ is computed from the activation delay measurement. The fitting parameters $\Isupp$ and $\Jideal(\ih=0)$, together with the plateau level $\Jconstr$ are extracted from switching current density ($\Jsw(\ih)$) curves. \textbf{(c)} The fitting parameters are further fed into SPICE, computing the nanowire temperature, switching current, and retrapping current at each simulation time step thanks to the corresponding fitting functions. The heat transfer through the oxide layer is simulated by a lumped-element RC circuit (represented here with the function $f(\ih(t), \tauth)$, giving rise to a delay in the hTron response to an input heater pulse. \textbf{(d)} The switching and retrapping current are finally used to define the hotspot behavior, that is either growing or retrapping depending on the current flowing in the channel $I_\mathrm{Ch}$. The device geometry and NbN resistance define the hotspot growth rate and normal resistance.}
    \label{fig:methods_model}
\end{figure*}

\subsection{SPICE Implementation}

Next, we describe the steps undertaken to build the simulation model in SPICE \cite{spice_book}, thus enabling the use of the modeled hTron device in circuits. The SPICE model file is available in the Supplemental Material.

As our simulation parameters were extracted from measurement data, a given device --- channel width $\wc$ and heater width $\wh$ --- had to be characterized prior to simulation. The NbN thicknesses were measured using an ellipsometer \cite{medeiros2019measuring}, and the widths and lengths of the channel and heater were measured from scanning electron micrographs. The NbN sheet resistance, which was used in the hotspot growth model, was also acquired. The steps showing the implementation of a SPICE model from the electrical characterization of a device are schematically represented in Figure \ref{fig:methods_model}. We next describe this procedure step by step.

In a first step (Figure \ref{fig:methods_model}a), electrical measurements must be performed. The activation delay $\tauon$ (delay between the input of a heater pulse and the channel switching) is measured to further define the transient behavior of the device. The details of this measurement will be discussed in Section \ref{sec:Section3}. The heater current-dependent switching current density $\Jsw(\ih)$ is also measured to define the static device behavior.

Subsequently (Figure \ref{fig:methods_model}b), the time constant of the heat transfer sub-circuit $\tauth$ is directly computed from $\tauon$ in SPICE, with the method described in Appendix $\ref{appendix:derive_heat_transfer}$. The suppressing current $\Isupp$, the unconstricted switching current at substrate temperature $\Jideal(\ih=0)$, and the plateau level $\Jconstr$ are extracted from $\Jsw(\ih)$ curves using the method detailed in Section \ref{sec:Section3}. These parameters are crucial to define the SPICE behavioral model. 

In the third step (Figure \ref{fig:methods_model}c), the SPICE behavioral model uses the fitting parameters to compute the switching current $\Isw$ and retrapping current $\Ir$ from a heater current input. The temporal behavior of the heat transfer between the heater and the NbN layer through the oxide is modeled using a simple heat transfer circuit with lumped elements (low-pass filter), for which the time constant is $RC=\tauth$. Without this added sub-circuit --- with output node denoted $V_\mathrm{heat}$ --- the temperature of the nanowire would start to grow instantaneously with a heater current input, which does not reflect real observations. The channel temperature and switching current are then computed using expression-based functions defined using the \texttt{.func} SPICE directive. This directive computes the value of a custom expression without the need to add a voltage node, and a SPICE behavioral source further allows one to probe the value of the function at each time step. Due to the low computational cost and analytical nature of the fitting functions, the SPICE implementation of our fitting functions is a straightforward process. This simplicity is required for a fast computation speed, as the solver updates the channel temperature, switching current and retrapping current at each time step, in response to the input heater current.

Finally, the channel current is compared to the switching and retrapping currents to compute the hotspot resistance (Figure \ref{fig:methods_model}d). A channel current above the switching current threshold would induce hotspot growth. The hotspot will eventually reset by decreasing the channel current below the retrapping current. The SPICE model of the non-linear inductor and hotspot behavior was based on the superconducting nanowire SPICE model introduced by Berggren et al. \cite{berggren2018superconducting}. Following \cite{torque2023}, the model was modified to embed a hotspot growth circuit based on the built-in SPICE integrator function \texttt{sdt()}, which improves convergence of the model and reduces the number of timesteps.


\section{Results}
\label{sec:Section3}

\begin{figure*}[t]
    \centering
    \includegraphics[width=\textwidth]{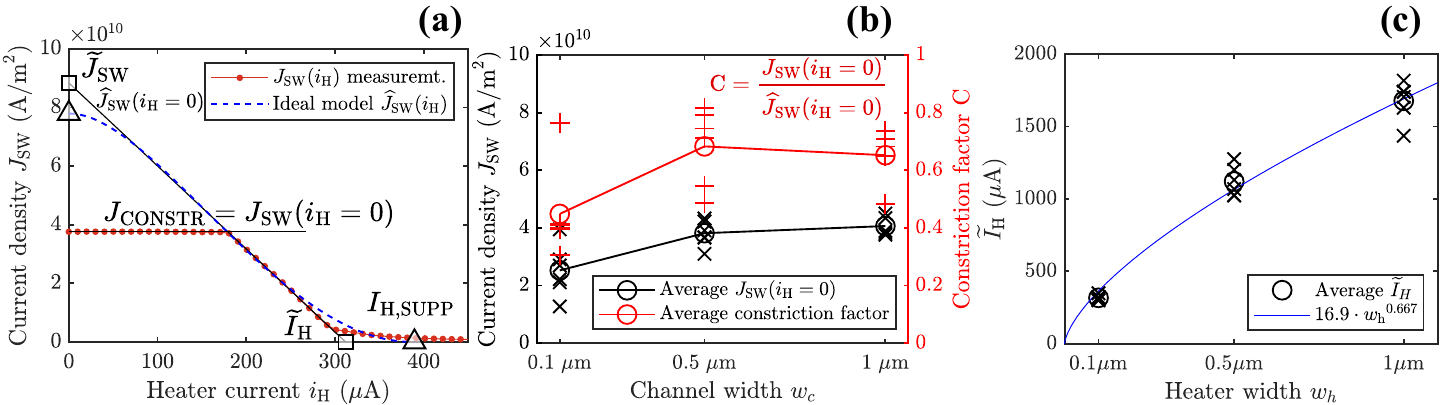}
    \caption{\textbf{hTron model design, and prediction of the fitting parameters from its geometry. }\textbf{(a)} Plot of switching current density against heater current highlighting our method developed to extract the model parameters from $\Jsw(\ih)$ measurements. A line is fitted to the linear part of the measurement data, which sets the fitting parameters $\Ihtilde$ and $\Jswtilde$ (indicated by $\Box$). These parameters have no simple physical meaning. The suppressing current $\Isupp$ and the unconstricted switching current at substrate temperature $\Jideal(\ih=0)$ (indicated by $\triangle$), that define the unconstricted model $\Jideal(\ih)$ can be further derived using $\Ihtilde$ and $\Jswtilde$. In this example, a constriction strongly limits the switching current of the measured device at low heater current, but it can be included in the model by defining $\Jmodel(\ih)=\mathrm{min}\{\Jideal(\ih),\Jconstr\}$, as seen earlier in Section \ref{sec:Section2}. \textbf{(b)} In black: measured switching current density at substrate temperature $\Jsw(\ih=0)$ against the device channel width. In red: constriction factor $C$, defined as the measured critical current density divided by the predicted switching current density. The average constriction factor is similar for $\wc=\SI{1}{\um}$ and $\SI{0.5}{\um}$ but decreases significantly for $\wc=\SI{0.1}{\um}$. \textbf{(c)} Extracted fitting parameter $\Ihtilde$ dependence on the heater width, with a power-law fit. The model suppressing current $\Isupp$ can thus be predicted by knowing the device heater width only. For plots \textbf{(b)} and \textbf{(c)}, each average is done over the six devices sharing the indicated heater or channel width. The measurement data points used to compute the averages are indicated with red $+$ markers or black $\times$ markers for clarity.}
    \label{fig:results_measurements}
\end{figure*}

In this chapter, we present the results of our research, fabrication, device characterization, and model design. We further analyze the model parameters and explain their dependence on device geometry. Finally, we compare the simulations and measurements of the device activation delay.

\subsection{Fabrication results}

The deposited NbN film shows a critical temperature of $\Tc\simeq\SI{12.5}{\K}$, a thickness --- measured with ellipsometry --- of $d_\text{c}=\SI{23.6}{\nm}$, and a sheet resistance of $\Rsheet=\SI{77.9}{\ohm/}\square$. The length of the channel and heater strips is  $l_\text{c}=l_\text{h}=\SI{10}{\um}$. Using these parameters, a depairing current of $J_\text{dep}=\SI{16.44e10}{\A/\m\squared}$ was computed using the Equation 13 of Charaev et al \cite{charaev2017proximity}. After the chip was cooled down to $\Tsub$, heater resistances of $\SI{44.2}{\ohm}$, $\SI{10.4}{\ohm}$, and $\SI{7.2}{\ohm}$ were measured on devices with heater widths of $\SI{0.1}{\um}$, $\SI{0.5}{\um}$, and $\SI{1}{\um}$ respectively. However, variations were observed among devices with the same geometry. The heater resistance could be engineered by tuning the heater length, which would not impair the device's operation. However, heating power is wasted on the substrate if the heater length exceeds the channel width.

\subsection{Device characterization}

Figure \ref{fig:results_measurements}a shows the dependence of the measured switching current density on heater current $\Jsw(\ih)$ (red circles). These data were used to extract the two fitting parameters for each device, $\Jideal(\ih=0)$ and $\Isupp$, together with the additional value $\Jconstr$. At low heater currents, the switching current remained close to constant, which we call the plateau (see Section \ref{subsec:constriction} for more details). After a sharp elbow, the plateau collapsed, and the switching current decreased linearly until it reached zero when the heater current equaled the suppressing current, $\ih=\Isupp$. As we explain in Section \ref{subsec:Isupp}, the suppressing current was correlated to the heater width. The full characterization of the measured devices can be found in the Supplemental Material.

 \subsection{Parameter Extraction}
 
Here we explain the developed step-by-step method to extract the two model parameters $\Jideal(\ih=0)$ and $\Isupp$ from a particular $\Jsw(\ih)$ measurement curve, represented schematically with a blue dashed line on Figure \ref{fig:results_measurements}a. The key idea is to recover the switching current that would have been observed if the constriction outside the heated area (\textit{i.e}., the plateau) had not been there. To do so, we used the information contained in the linear part of the $\Jsw(\ih)$ measurement data: 
\begin{enumerate}
    \item We fit a straight line to the linear part of the curve, ignoring the plateau on the left side and the part on the right side where the switching current reaches zero.
    \item This linear fit defines $\Jswtilde$ and $\Ihtilde$, the intersection between the line and the y and x axis, respectively.
    \item The fitting parameters are then recovered : $\Jideal(\ih=0)= \alpha\cdot\Jswtilde$, and $\Isupp= \beta\cdot\Ihtilde$. The $\alpha$ and $\beta$ constants are correction parameters obtained by fitting a straight line to the fitting function directly. These constants simply account for the fitting function's curvature and are valid for the same value of $\eta$ (see the Supplemental Material for more information). We found $\alpha=0.88$ and $\beta=1.25$ for $\eta=2$. Plugging $\Isupp$ into Equation \ref{eq:T_vs_Ih} and $\Jideal(\ih=0)$ into Equation \ref{eq:J_vs_Tnw} give the unconstricted model $\Jideal(\ih)$ as defined in Equation \ref{eq:Jideal}, which represents the measurement we would observe if there was no plateau.
    \item The final model expression --- including the plateau --- is defined in Equation \ref{eq:Jmodel} as the minimum between the unconstricted model $\Jideal(\ih)$ and the constriction level $\Jconstr$.
\end{enumerate}

It is noteworthy that any $\Jconstr$ value can be set in the model to test the effect of various constriction levels in a given circuit.

\subsection{Effects of constrictions outside of the heated area}
\label{subsec:constriction}
 
The presence of a plateau at low heater current in Figure \ref{fig:results_measurements}a suggests that the measured device switching current was limited by one or multiple weak spots located along the channel, away from the heated area. At low heater currents, these weak spots switched at lower channel currents than the channel portion below the heater. We therefore observed a lower switching current than that of the heated region. As the heater current increased, a point was reached where the heated part of the channel started to switch at a current lower than that of the weak spot. This marked a significant change, as the switching current was now modulated by the heater current, and the plateau thus abruptly collapsed. Beyond this point, with further increase in heater current, the switching occurred directly under the heater, until the channel was completely suppressed. This plateau was also observed by Baghdadi et. al. \cite{baghdadi2020multilayered}.

Constrictions are common candidates for weak spots in superconducting nanowires \cite{kerman2007constriction}. In Figure \ref{fig:results_measurements}b (in black), we plot the average switching current density of the weak spot $\Jsw(\ih=0)$ against the channel width, for all 17 measured devices (see Supplemental Material for all measurement curves). The switching current density of the weak spot was similar for $\SI{1}{\um}$-wide channels and $\SI{0.5}{\um}$-wide channels, but decreased significantly for $\SI{0.1}{\um}$-wide channels, while also showing greater variance. The same trend can also be seen with the constriction factor $C$, plotted in blue, defined as the ratio between the measured switching current at substrate temperature $\Jsw(\ih=0)$ and the ideal predicted switching current at substrate temperature $\Jideal(\ih=0)$ \cite{frasca2019determining}. The fact that the devices with the narrowest channels were more constricted may have resulted from the channel line-width roughness, which may have a greater impact on the switching current as the width decreases \cite{colangelo2022large}.

Finally, if our fabrication process was perfect, $\Jideal(\ih=0)$ should, in theory, be the same for all devices within the same wafer, and should not depend on the channel width (see Appendix \ref{appendix:minimal_set_parameters} for more details). In practice, despite some variance, the average $\Jideal(\ih=0)$ value was indeed comparable across devices with different channel widths, implying that the channel's average switching current density was similar among devices, despite showing localized constrictions.
 
\subsection{Analysis and prediction of the suppressing current}
\label{subsec:Isupp}

The suppressing current parameter $\Isupp$ had smaller values for narrower heaters, and could be expressed as a function of the heater width $\wh$ quite accurately, as shown in Figure \ref{fig:results_measurements}c. As seen in the Supplemental Material, the suppressing current was not correlated with the channel width, which implies that devices with the same heater width but different channel width would thus be fully suppressed at the same heater current. The function $\Ihtilde(\wh)=16.9\cdot\wh^{0.667}$, resulting from a power-law fitting, predicts the suppressing current for different heater widths. As reducing $\wh$ increases both the resistance and the dissipated power per unit area, the suppressing current decreases faster for narrower heaters. Here, the fitting parameter of the power-law function, valid for an entire wafer, includes information about the oxide thickness and heater's material properties. This general formula allows us to estimate the behavior of devices with heater widths other than those specifically measured.

\begin{figure*}[t]
    \includegraphics[width=\textwidth]{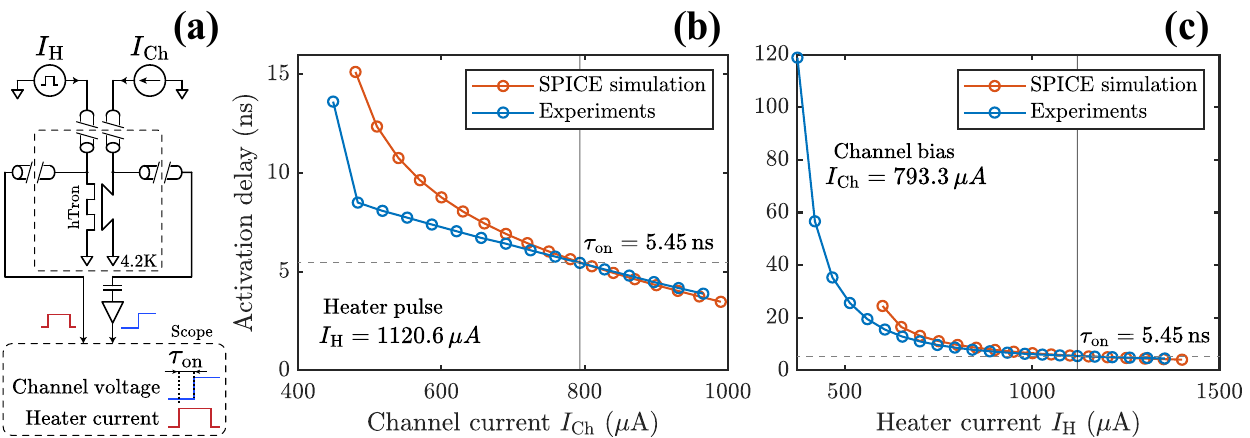}
    \caption{\textbf{Prediction of the hTron maximum operating speed.} \textbf{(a)} Simplified schematics of the measurement setup of the hTron activation delay $\tauon$, defined as the channel response time following the application of a heater current pulse of magnitude $\Ih$. The heater current pulse duration is much larger than $\tauon$. We plotted the measured and simulated activation delay against \textbf{(b)} the channel bias current (for a constant heater current pulse amplitude), and \textbf{(c)} the heater current amplitude (for a constant channel bias current). The graphs highlight the model's predictive capability across a large span of input currents, despite a deviation from actual measurements at lower currents. This device channel width and heater width are $\SI{1}{\um}$ and $\SI{0.5}{\um}$, respectively.}
    \label{fig:results_delay}
\end{figure*}

\subsection{hTron maximum operating speed}

The transient response of the hTron is mainly defined by the heat transfer from the heater to the channel through the oxide. It limits the operating speed of the hTron devices, resulting in a non-zero device activation delay, $\tauon$. This delay, measured as the time needed to observe a channel switch after the input of a heater pulse, is a key aspect of the device behavior.

In Figure \ref{fig:results_delay}a, we show the setup used to measure the activation delay $\tauon$ at a specific $(\Ih,I_\text{Ch})$ operating point. The results of sweeping the channel current while maintaining a fixed heater current are shown in Figure \ref{fig:results_delay}b. Figure \ref{fig:results_delay}c illustrates the results when the heater current is varied with a fixed channel current. In order to reproduce this data in SPICE, we first chose an operating point $(\Ih=\SI{1120.6}{\uA},I_\text{Ch}=\SI{793.3}{\uA})$, and extracted the activation delay from the experiments ($\tauon=\SI{5.45}{\nano\s}$). We further obtained the RC time constant of our heat transfer sub-circuit of $RC=\tauth=\SI{3.86}{\ns}$ directly from the activation delay value $\tauon$ (details can be found in Appendix \ref{appendix:derive_heat_transfer}). Sweeping the input currents in the same way in SPICE as we did in the experiments, we observed that the model was reliable around the operating point $(\Ih=\SI{1120.6}{\uA},I_\text{Ch}=\SI{793.3}{\uA})$ and for currents higher than the bias point.

This simple and efficient method allowed us to simulate the temporal behavior of the hTron device and can be used to determine the maximum operating speed in real circuits. As seen in Figure \ref{fig:results_delay}b and \ref{fig:results_delay}c, there is a clear disagreement between the model and experiment at low bias currents. This disagreement is possibly due to the variation between the modeled and actual switching current, which we overestimated in this example. Indeed, we can see that the simulated delay is greater than the measured delay at low bias currents, meaning that the device switched \textit{earlier} in experiments than in simulation. Even a small difference in switching current can make a large deviation in the predicted delay.

\section{Assessment of model performance}
\label{sec:Section4}

Finally, we evaluated our model's execution speed by running the previously published simplified 0D electrothermal model in SPICE and comparing it to our model in a $\SI{250}{\nano\s}$-long simulation of a shunted channel (see Supplemental Material for the complete analysis). The simplified 0D electrothermal model, as implemented by Castellani \cite{castellani2020thesis}, solves heat equations with SPICE behavioral sources. To compare both models, we have set the SPICE parameter $\text{\texttt{reltol}}=10^{-6}$ for both models, as suggested by El Dandachi \cite{torque2023}, and gathered the simulation results in Table \ref{tab:results}. In a first simulation run, we varied the maximum simulation time step $\Delta t_\mathrm{max}$ from $\SI{1}{\pico\s}$ to $\SI{10}{\pico\s}$ over 11 steps, and our model took less than eleven seconds to run, while the previous model took eight hours. In a second run, we varied the maximum time step from $\SI{1}{\femto\s}$ to $\SI{50}{\pico\s}$ across 48 steps, and our model took less than four hours to run, while the previous model had to be halted after two days because it was stuck at a significantly slow simulation speed in the order of $\approx\SI{1}{\femto\second/\second}$.

To conclude, the enhanced simulation speed of our model can be mainly explained by two factors. First, the previous model required solving a greater number of nodes due to its approach of using behavioral sources to solve the heat transfer equations of the entire system. Second, unlike our model, it did not incorporate the enhanced hotspot integrator sub-circuit, potentially leading to a higher number of time points to solve for \cite{torque2023}. All our simulations were conducted on the same personal computer, highlighting the difference in simulation speed between the previous and new models. We successfully reduced typical simulation times of the hTron device from hours to mere minutes or even seconds, making quick simulations feasible in practice on a personal computer for the first time.

\begin{table*}[]
\begin{tabular}{l|ll|ll}
\multirow{2}{*}{} & \multicolumn{2}{c|}{\textbf{Run 1}: $\Delta t_\text{max}=\SI{1}{\pico\s}$...$\SI{10}{\pico\s}$}                               & \multicolumn{2}{c}{\textbf{Run 2}: $\Delta t_\text{max} = \SI{1}{\femto\s}$...$\SI{50}{\pico\s}$}                            \\
                       & \multicolumn{1}{l|}{\# Time points}  & Simulation Time      & \multicolumn{1}{l|}{\# Time points} & Simulation Time   \\ \hline
Previous model         & \multicolumn{1}{l|}{813,628} & $\num{8.0}$ hours   & \multicolumn{1}{l|}{N/A}            & N/A               \\
Our model             & \multicolumn{1}{l|}{2,045}    & $\num{10.8}$ seconds & \multicolumn{1}{l|}{1,189,652}    & $\num{3.3}$ hours
\end{tabular}
\caption{Comparison of simulation speed between the previous model designed by Baghdadi et al. \cite{baghdadi2020multilayered} and our model for the same SPICE circuit but different time resolutions.}
\label{tab:results}
\end{table*}


\section{Discussion}
\label{sec:Section5}

This section discusses the challenges of predicting weak spots in superconducting nanowires, the impact of hidden parameters in our model, and the prediction of the device activation delay and operating speed, highlighting both the strengths and limitations of our approach.

Locating and removing constrictions in superconducting nanowires is essential both in photon detection and hTron device operation. However, in the current state of research, accurately predicting the location and number of weak spots --- or areas with reduced critical current --- along a channel remains unfeasible. Weak spots located away from the heated area are responsible for the plateau. They are unwanted and are not part of the normal device behavior. Therefore, we ignored the plateau during the parameter extraction, fitting only to the linear part of the $\Jsw(\ih)$ plots. As a result, the ideal model $\Jideal(\ih)$ was a hypothetical curve that probed the critical current directly below the heater, theoretically representing the channel's critical current as if there were no weak spots. The plateau could then be set using the constriction level $\Jconstr$. While we observed that the plateau level was smaller for narrower channels, it could not be accurately predicted due to its intrinsic stochastic nature. 

In this study, the measured devices were from a single wafer, thus some geometrical parameters such as the oxide thickness did not explicitly appear in the fitting equations and were hidden. Consequently, new measurements and recalibration of the model's parameters may be required to predict the behavior of devices with a different oxide thickness. Even though the previous model did not present hidden parameters, its prediction capabilities were limited as well. Indeed, it contained various thermal parameters that were either extracted from literature or fitted from measurements, and could not be predicted from geometry. In contrast, our high-level curve-fitting approach simplifies the analysis, allowing for a systematic process of parameter extraction for any geometry, and outperforms the previous work in accuracy and speed.

Finally, the heat transfer from the heater to the channel through the oxide is a critical factor limiting the operating speed of the devices, resulting in a non-zero device activation delay, $\tauon$. In complex circuits, this delay becomes an increasingly significant figure for circuit designers. Indeed, accurately predicting and simulating the maximum operating speed of a real-world circuit is essential. In our approach, we intentionally over-simplified the way we modeled the heat transfer, offering the advantage of only including one parameter --- an RC time constant denoted $\tauth$, which can be directly obtained from $\tauon$ ---, making it straightforward to implement in practice. However, as seen in Figure \ref{fig:results_delay}, our model's prediction of the activation delay was less accurate for low input currents. At higher currents, though, our model accurately predicted the activation delay, allowing one to simulate the temporal behavior of a hTron-based circuit close to the maximum operating speed. As a final note, while the minimum measured activation delay was in the order of $\qtyrange{1}{10}{\ns}$ for our particular device, it is noteworthy that the fabricated devices were not designed for speed. A thinner oxide layer, as well as narrower heater wires, would greatly improve the operating speed of hTron devices, as seen in \cite{baghdadi2020multilayered} where the oxide layer is thinner.

As a final note, even though we used the LTspice framework, our model uses generic SPICE functions that can be translated to other simulators. The stability of our model in these other simulators was however not tested, and some special functions such as \texttt{sdt()} may not be available in every SPICE implementation. However, the equivalent circuit integrator from the original nanowire model can be used \cite{berggren2018superconducting}.


\section{Conclusion}

Our approach to modeling the hTron device relies on the fitting of simple physics-informed equations to a relatively large amount of experimental data. All the fitting parameters are extracted from basic measurements, and no arbitrary parameter has to be tuned to better fit our measurements. The static behavior was modeled from the critical current measurements performed on 17 hTrons having 9 different geometries. We were able to extract the two relevant fitting parameters and extrapolate them for hTron geometries that were not measured. Moreover, we were able to simulate the hTron transient response over a large span of current inputs thanks to activation delay measurements --- delay between the application of a heater current pulse and the switching of the channel. This transient behavior is critical in circuit design as it sets the maximum operating speed of a circuit. We applied our model to measurement data published by Baghdadi et al. \cite{baghdadi2020multilayered}, and obtained a better agreement than their proposed model. Moreover, we compared the simulation speed of both approaches on a personal computer in SPICE, and obtained a result in seconds, while the previous approach completed the same simulation in hours. Consequently, our model and SPICE implementation are tailored for efficient design iterations in nanowire-based electronics. Finally, our model effectively simulates devices for any constriction --- or plateau --- level, a critical feature given the unpredictability of these occurrences.

Our findings corroborate the positioning of the hTron as an alternative to the nTron device for applications requiring both high output impedance and electrical isolation between the gate and the channel. From a broader viewpoint, superconducting nanowires could complement Josephson Junctions in areas where they are currently lacking. Looking forward, we will use our model to solve larger and more complex circuits like SNSPD array readouts or memory arrays. The high simulation speed capability, accuracy, and simplicity of our approach positions this work as a promising first step for the building of useful simulation tools for nanowire-based circuits.

\section*{Acknowledgements}
The initial stages of the research were sponsored by the U.S. Department of Energy, Office of Science, Office of Basic Energy Sciences, under Award Number DE-AC02-07CH11359. The completion of data analysis and manuscript preparation were sponsored by the National Science Foundation under Grant No. OMA-2137723. The authors thank Alessandro Restelli, Joshua Bienfang and Ilya Charaev for helpful scientific discussions. V.K. would like to thank Edoardo Charbon from the Advanced Quantum Architecture (AQUA) Laboratory at the Swiss Federal Institute of Technology (EPFL). O.M. acknowledges support from the NDSEG Fellowship program. M. Colangelo acknowledges support from the MIT Claude E. Shannon award.

\begin{appendices}

\section{Derivation of the heat transfer parameter}
\label{appendix:derive_heat_transfer}

As shown in Figure \ref{fig:methods_model}, the measured activation delay $\tauon$ at a defined bias point $(\Ih,I_\text{Ch})$ is used to compute a model parameter, $\tauth$, which sets a delay in the transient response of the device. This delay is approximated in SPICE by feeding the heater current into a low-pass filter before computing the switching current, mimicking a delay in the response of the device.

First, by plugging the channel bias current $I_\text{Ch}$ into  Equation \ref{eq:J_vs_Tnw} and solving for the channel temperature, we can define the needed channel temperature to make the channel switch at this particular bias. We call this temperature the switching temperature $T_\mathrm{SW}$: 

\begin{equation} \label{eq:Tswitch}
T_\mathrm{SW}=\Tc \cdot  \left( 1 - \left( \frac{I_\text{Ch}}{\Jc \cdot d_\text{c} \cdot \wc} \right)^{\frac{1}{2.1}} \right)^{\frac{1}{3}}
\end{equation}

By solving for the heater current in Equation \ref{eq:T_vs_Ih}, we then find the heater current needed to reach this switching temperature, called $I_\mathrm{H,SW}$ is given by:

\begin{equation} \label{eq:Ih_switch}
I_\mathrm{H,SW} = \left( \frac{T_\mathrm{SW}^4 - \Tsub^4}{\Tc^4 - \Tsub^4} \right)^{\frac{1}{\eta}} \cdot \Isupp
\end{equation}

The voltage at the filter output node --- called \texttt{heat} in our SPICE model --- is $V_{\mathrm{heat}}(t)=\ih(t)\cdot(1-e^{-t/\tauth})$, where $\ih(t)$ is the device input heater current, and $\tau_{\mathrm{filter}}=RC$ is our unknown. We can thus finally find the $RC$ time constant that makes $V_{\mathrm{heat}}(t)$ reach $I_\mathrm{H,SW}$ in time $t=\tauon$:

\begin{equation} \label{eq:tau_filter}
\tauth= -\frac{\tauon}{\ln\left(1 - \frac{I_\mathrm{H,SW}}{\Ih}\right)}
\end{equation}

We set $R=\SI{1}{\ohm}$, and $C=\tauth$.

\section{Minimal set of parameters required to model the hTron static behavior}
\label{appendix:minimal_set_parameters}

\begin{figure*}[t]
    \centering
    \includegraphics[width=0.8\textwidth]{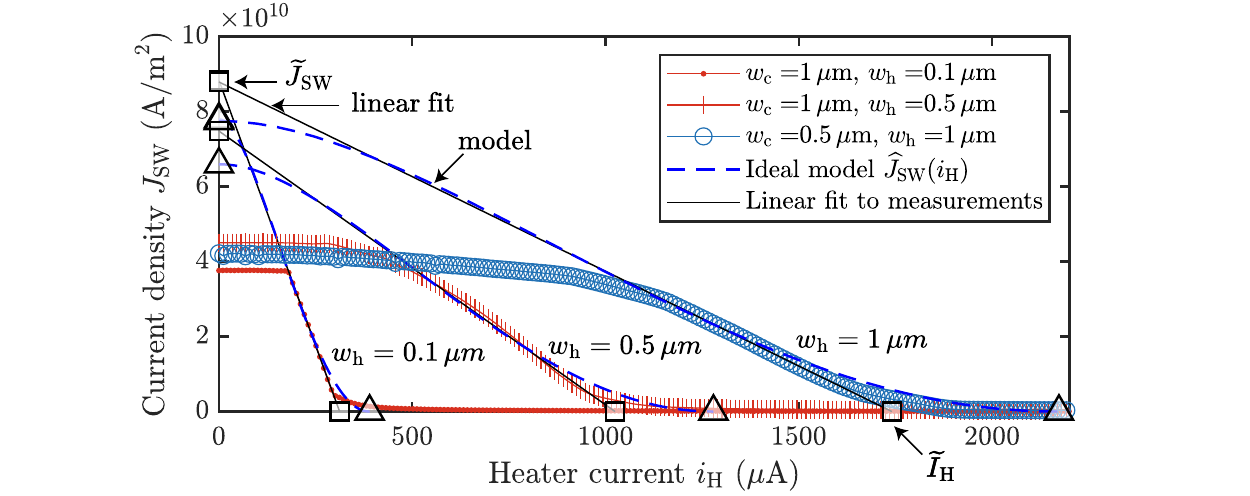}
    \caption{\textbf{Static behavior modeling of three devices with different heater widths.} These measurement plots of switching current against heater current show that the linear part of the  $\Jsw(\ih)$ curves tends to have a common origin at the y-axis ($\square$ markers on the y-axis).} 
    \label{fig:appendix_results_models}
\end{figure*}

As we saw throughout this paper (see Figure \ref{fig:results_measurements}a), the static behavior of an hTron device can be represented by two fitting parameters only: the ideal switching current density at substrate temperature $\Jideal(\ih=0)$ (computed from the fitting parameter $\Jswtilde$) and the suppressing current $\Isupp$ (computed from the fitting parameter $\Ihtilde$). The constricted switching current density $\Jconstr$ can be set greater than $\Jideal(\ih=0)$ to simulate an unconstricted device, or set arbitrarily. 

Figure \ref{fig:appendix_results_models} shows the measurement data, linear fit, and the model curve (dotted blue line) for three different devices with three different heater widths and arbitrary channel widths. The Figure shows explicitly that all $\Jsw(\ih)$ curves tend to share the same $\Jswtilde$ parameter, whatever the device geometry. However, variations and imperfections between devices are expected, leading to variations in the $\Jswtilde$ parameter, thus in the predicted $\Jideal(\ih=0)$. As seen earlier (see Figure \ref{fig:results_measurements}c) the parameter $\Ihtilde$ can be approximated from the heater width with a power-law fit, which is also valid for an entire wafer. Once this fit is found, the heater width and the $\Jswtilde$ parameter are, in theory, the only two required parameters to model any hTron geometry on a given wafer.

\end{appendices}

\newpage
\bibliography{main}

\end{document}


\maketitle
\vspace{6pt}

\date{June 28, 2024}

\section{Characterization of the measured devices}
\label{appendix:characterize_devices}

\subsection{Characterization of the hTron static behavior: Switching current measurements}

\begin{figure}[h]
    \centering
    \includegraphics[width=\textwidth]{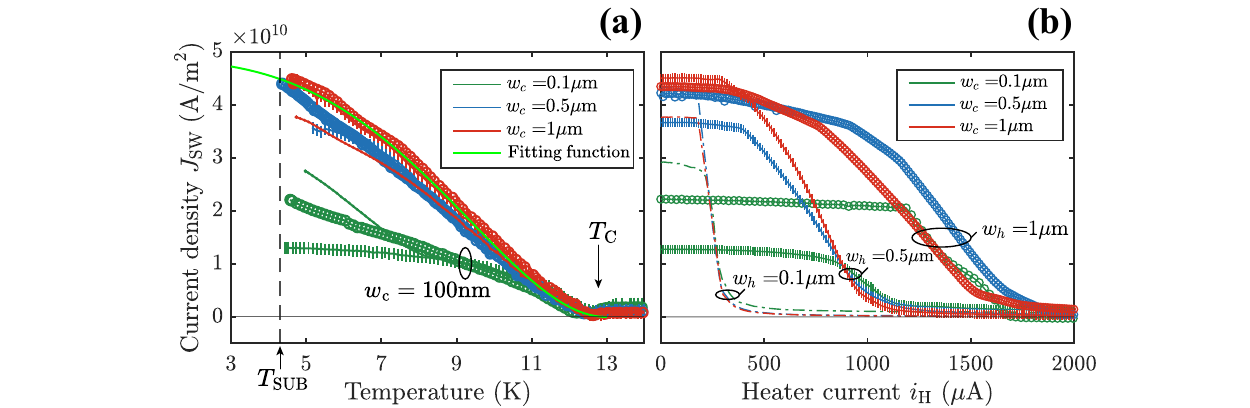}
    \caption{\textbf{Measurement results and method for extracting model parameters}\\ Experimental data of the switching current density of 9 different hTron geometries, plotted against \textbf{(a)} the substrate temperature and against \textbf{(b)} the heater current. Colors represent channel widths and markers heater widths. The function $\Jsw(T) = \num{4.48e10}\cdot\left(1-\left(T/\Tc\right)^3\right)^{2.1}[\unit{\A/\m\squared}]$ was used to fit to the data curve of the device with $\wh=\SI{0.5}{\um}$ and $\wc=\SI{1}{\um}$ (red curve with vertical markers), showing a smooth transition from $\Tsub$ to $\Tc$. Heater-dependent switching current curves show a plateau at low heater currents, and a clear separation by heater widths: narrower heaters require less current to suppress the channels.}
    \label{fig:results_measurements}
\end{figure}

The complete measurement set of 9 devices with different geometries is shown on Figure \ref{fig:results_measurements}a and Figure \ref{fig:results_measurements}b. In Figure \ref{fig:results_measurements}a, one can see that the channel switching current density of measured devices is smoothly reduced with temperature until the channel is fully suppressed at $\Tc\simeq\SI{12.5}{\K}$. Moreover, the experimental data agrees nicely with the fitting function $\Jsw(T) = \num{4.48e10}\cdot\left(1-\left(T/\Tc\right)^3\right)^{2.1}[\unit{\A/\m\squared}]$, that was fitted to the device with $\wh=\SI{0.5}{\um}$ and $\wc=\SI{1}{\um}$ (red curve with vertical markers). All other measurement curves can be fitted independantly. On the other hand, Figure \ref{fig:results_measurements}b shows the switching current density's dependence on heater current. This plot is used to extract the two fitting parameters $\Jideal(\ih=0)$ and $\Isupp$, together with the additional parameter $\Jconstr$.

Two comments can be made on these plots : (1) a plateau appears for low heater currents for almost all devices, with a much stronger suppression for narrower channels (green curves) and (2) the suppressing current $\Isupp$ is correlated to the heater width but not to the channel width, as all devices with same heater width are fully suppressed at the same heater current. \\

\subsection{Characterization of the hTron transient behavior: activation delay measurements}

\begin{figure}[h]
    \centering
    \includegraphics[width=\textwidth]{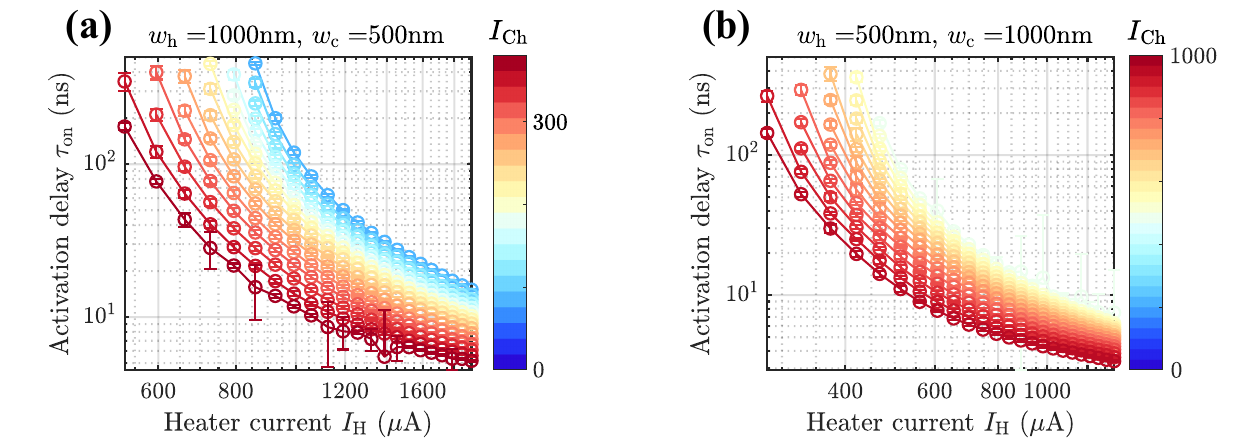}
    \caption{\textbf{Activation delay measurements of two devices with different geometries.} Device \textbf{(a)} has a heater width of $\SI{1000}{\nm}$ and channel width of $\SI{500}{\nm}$, and \textbf{(b)} has a heater width of $\SI{500}{\nm}$ and channel width of $\SI{1000}{\nm}$. The measurement technique is identical to the one described in the paper. The log-log scale helps differentiate the lines from each other and highlights the exponential nature of the curves. The activation delay decreases with the heater and channel current and reaches a plateau on the bottom-right part of the plot. More channel or heater current would indeed overcome the switching current. The average and standard deviation over 100 trials are displayed. A larger variance is observed for lower heater and channel currents, showing the stochastic nature of the heat transfer coupled with the switching dynamics of the channel. A few points have an exceptionally high variance because of some outliers in the measurement process.}
    \label{fig:supplement_delay}
\end{figure}

In this subsection, we comment on the activation delay measurements performed on two devices, displayed on Figure \ref{fig:supplement_delay}.

For every channel current bias, different heater pulse amplitudes were sent to the heater. These plots represent the dependence of the delay with the channel bias and heater pulse amplitude. Figure 4 of our paper shows the impedance-matched experimental setup used to preserve fast rising edges while avoiding signal reflections.

The exact time delay between the arrival of the heater pulse and the time at which the channel switches (the activation delay) allowed us to quantify the hTron response time to a heater pulse, which limits the device operational speed. The Figure \ref{fig:supplement_delay} shows this activation delay for two devices with different heater and channel widths across their full range of heater and channel currents. The channel bias current increases from blue (low) to red (high current), and the heater pulse current amplitude increases from left to right. The vertical asymptote on the left-hand side gives the smallest heater pulse that leads to a switch at a particular channel bias, a smaller channel bias requiring more heat to switch the channel. The horizontal asymptote on the right-hand side of the plot gives the smallest achievable response time. For higher heater currents, the suppressing current is reached, and for higher channel currents, the switching current is reached, which makes the channel continuously normal in either case.

The measured device with a narrower heater (Figure \ref{fig:supplement_delay}b) transfers the heat through the oxide layer in a more efficient way than the device with a wider heater (Figure \ref{fig:supplement_delay}a): the minimum reached activation delay is substantially lower at maximum heater current and maximum channel current --- respectively, close to the suppressing current and to the switching current. This is primarily due to the fact that a narrower heater localizes the heat on a more focused spot of the channel, leading to a quicker increase in temperature on the channel region located directly above the heater. A higher heater resistance is thought to play a role in getting a faster activation delay as well.

\section{Reproduction of previously published results}
\label{appendix:reproduce_baghdadi}

\begin{figure}[t]
    \centering
    \includegraphics[width=\textwidth]{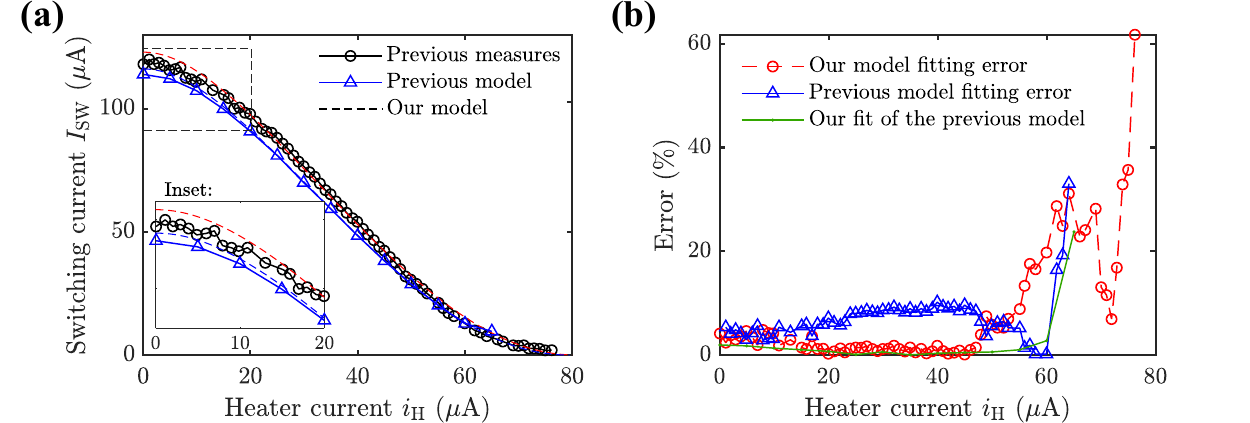}
    \caption{\textbf{Comparison between our static model and the previously published model. }\textbf{(a)} Previously published switching current measurements (black circles) and model (blue triangles) extracted from \cite{baghdadi2020multilayered}. The dotted red and blue curves are our predictions for their measurements and their model, respectively. We used the method introduced in our paper to extract our modeling parameters from these curves. \textbf{(b)} Error between the measurement data and our prediction (red), the measurement data and the prediction from \cite{baghdadi2020multilayered} (blue), and our fit to the simulation model from \cite{baghdadi2020multilayered} (green). Our model predicts their measurements, as well as their model with less than $5\%$ error on most of the heater current span, until around $\SI{50}{\uA}$.}
    \label{fig:results_baghdadi_model}
\end{figure}

\begin{figure*}[h]
    \includegraphics[width=\textwidth]{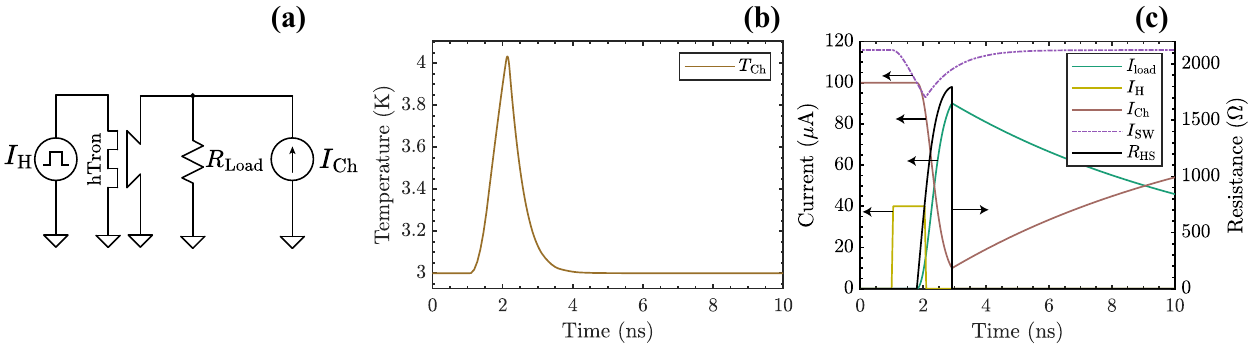}
    \caption{\textbf{Reproduction of Baghdadi et al. \cite{baghdadi2020multilayered} results from an $\Isw(\ih)$ measurement.} \textbf{(a)} Circuit used to simulate the hTron behavior in SPICE, identical to the one used in Figure 3 of \cite{baghdadi2020multilayered}. The power dissipated in the heater increases the channel's temperature $\TCh$, decreasing the channel switching current $\Isw$. As the channel switches to the normal phase, the bias current $I_\text{Ch}$ is transferred to a $\SI{50}{\ohm}$ load $R_\text{Load}$, allowing the channel to reset back to the superconducting state. SPICE simulation result of the \textbf{(b)} effective channel temperature and of the \textbf{(c)} electrical behavior of the hTron device. The channel temperature and switching current were computed using Equation 1 and Equation 2, respectively. The device and simulation parameters were given by \cite{baghdadi2020multilayered}, while the model parameters $\Isupp$ and $\Jsw(\ih=0)$ were extracted from the $\Isw(\ih)$ plots using our method presented in this work.}
    \label{fig:results_simulation}
\end{figure*}

In this section, we explain how we successfully replicated the transient behavior seen in Figure 3 of \cite{baghdadi2020multilayered}, \textit{i.e.}, the hTron response to a heater current pulse. We fitted our model to the previously published measurement data, allowing us to further simulate the hTron behavior in SPICE using the exact same inputs that were used in the previous publication.

\subsection{Reproduction of heater-current-dependent switching current curves}

We first fitted our model to the measured data of the device switching current (Figure 5 of \cite{baghdadi2020multilayered}), which we reported in Figure \ref{fig:results_baghdadi_model}a. The measured device switching current at substrate temperature is approximately $\Jsw(\ih=0)\simeq\SI{120}{\uA}$, and the $\Jsw(\ih)$ curve does not present a plateau at low heater current, unlike our devices. Our model is shown in dotted red.
The error between our model and the measurement data is plotted in red in Figure \ref{fig:results_baghdadi_model}b.
Our model closely matches the measurements, with less than $\SI{3}{\%}$ error over almost all the heater current range $\qtyrange{10}{50}{\uA}$. On the other hand, the previous 3D electrothermal model error is about $\qtyrange{8}{10}{\%}$ error in the same range. For heater currents over $\SI{50}{\uA}$ --- getting closer to the suppressing current --- the error increases for both the previous and the new model. This is to be expected, as the signal-to-noise ratio decreases, and the concept of switching current becomes ambiguous at low currents.
From these previous measurements, we extracted the parameters: $\Isupp=\SI{80.9}{\uA}$ and $\Jideal(\ih=0)=\SI{122.9}{\uA}$. 

The quality of our fit shows the validity of our model on a completely different dataset (different acquisition methods, oxide and NbN thickness, hTron geometry, etc). Moreover, unlike for our measurement data, our model is accurate even at low heater currents, which validates the fact that our ideal model would be accurate if there was no plateau in our measurement data. 

We then applied our fitting method to the previous 3D electrothermal model directly. The goal here is to replicate the 3D electrothermal model using our fitting equations, allowing us to perform SPICE simulations with a model as close as possible to what was done in the publication by Baghdadi et al. The resulting curve can be seen in Figure \ref{fig:results_baghdadi_model}a (blue dotted) and the according error is plotted in Figure \ref{fig:results_baghdadi_model}b (in green). Our model can fit the previous 3D electrothermal model with very low error, below $\SI{3}{\%}$ up to a heater current of $\SI{60}{\uA}$. We found the fitting parameters $\Isupp=\SI{79.5}{\uA}$ and $\Jideal(\ih=0)=\SI{116.1}{\uA}$, close to the value of the switching current at zero heater current of the previous model, $\Jsw(\ih=0)=\SI{113.9}{\uA}$. We computed a retrapping current value of $\Ir=\SI{20.3}{\uA}$ using the relation given in \cite{berggren2018superconducting}, which is comparable to Baghdadi et al.’s measured value of $\Ir\approx\SI{21}{\uA}$.\\

\subsection{Reproduction of the hTron transient behavior}

In a second step, we use our SPICE model with the extracted fitting parameters from the previous section to reproduce the transient behavior from Figure 3 of \cite{baghdadi2020multilayered}.

In this circuit (Figure \ref{fig:results_simulation}a), the hTron is shunted by a $\SI{50}{\ohm}$ load and biased at $\SI{100}{\uA}$. It switches after a $1$-$\unit{\ns}$ heater current pulse is applied, diverting the current into the resistor. The channel then reverts to its superconducting state, allowing current return with a time constant determined by the channel inductance and load resistance. Our SPICE simulation accurately reproduces the previous result. The channel temperature (Figure \ref{fig:results_simulation}b) rises to around $\SI{4}{\K}$ with a typical exponential shape, which switches it as a channel bias current is applied. The hotspot resistance (Figure \ref{fig:results_simulation}c) increases with a sharp spike as the heater pulse is applied and suddenly collapses when the channel goes back to the superconducting phase. The main difference between our model and the previous 3D electrothermal model is that our model does not account for the temperature rise from Joule heating when the channel is in the normal state. However, this effect is approximately taken into account by our hotspot growth rate equation \cite{berggren2018superconducting}, which makes our simulation valid.

To simulate the device transient behavior, we matched the device activation time with the one observed in Figure 3c of \cite{baghdadi2020multilayered}, estimate as $\tauon\approx\SI{750}{ps}$. This gives an RC time constant of $\tau_\mathrm{filter}=\SI{1.6}{\ns}$ (See Appendix A for more details about the heat transfer parameter).

All simulation parameters used from \cite{baghdadi2020multilayered} are listed here: $\wc=\SI{600}{\nm}$, $\wh=\SI{500}{\nm}$, channel inductance $L_\text{Ch}=\SI{500}{\nH}$, $\Tsub=\SI{3}{\K}$, $\Tc=\SI{8.4}{\K}$, NbN thickness $d_\text{c}=\SI{20}{\nano\m}$, $\Rsheet=\SI{470}{\ohm/}\square$. Note that the heater's thickness and resistance, embedded in our fitting parameters, are unused by our model.\\

\section{Comparison of the model behavior with different $\eta$ values}

\begin{figure}[h]
    \centering
    \includegraphics[width=0.45\textwidth]{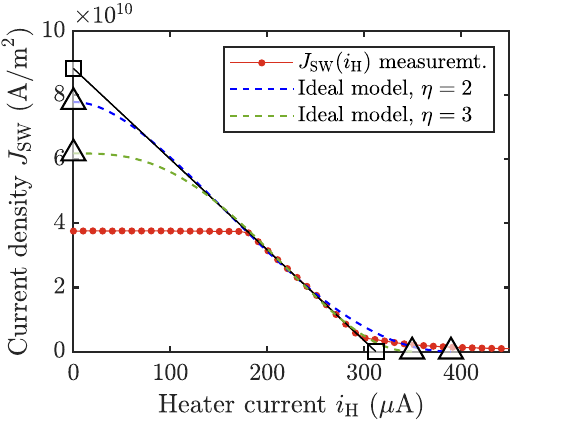}
    \caption{\textbf{Comparison between different eta values.} Measurement of switching current density against heater current, and ideal model for the two different values of $\eta$, $\eta=2$ and $\eta=3$. $\eta=3$ fits our measurements better, especially when the heater current gets close to the suppressing current ($\triangle$ marker on the x-axis). Note that $\eta=2$ is the value used throughout our paper (the blue dotted curve in this figure is thus the same as the blue dotted curve in Figure 3a of our paper).}
    \label{fig:results_different_eta}
\end{figure}

The model behaves differently with different $\eta$ values, used in Equation 1 of our paper, which represents the dependence of the channel temperature with the heater current. As seen in Figure \ref{fig:results_different_eta}, a value of $\eta=3$ aligns more closely with our measurement data, especially when $\ih$ is close to $\Isupp$. However, a value of $\eta=2$ agrees very nicely with Baghdadi et. al. measurement data, as seen in Figure \ref{fig:results_baghdadi_model}. We thus chose a value of $\eta=2$: our devices are severely constricted, showing a plateau in most cases, unlike devices measured in the Baghdadi et. al. paper. Thus more experimental data on less constricted devices would help determine the best value for $\eta$. While a more complicated equation could potentially better describe the channel temperature as a function of heater current, the strength of our approach lies in its simplicity: introducing an additional free parameter would counteract this benefit, making our model less straightforward without a substantial increase in predictive accuracy for our specific application.

Finally, while the value of the parameters $\Ihtilde$ and $\Jswtilde$ do not depend on the value of $\eta$, the $\Jideal(\ih=0)$ and $\Isupp$ parameters --- defined as $\Jideal(\ih=0)= \alpha\cdot\Jswtilde$, and $\Isupp= \beta\cdot\Ihtilde$ --- do. Indeed, $\Jideal(\ih=0)$ and $\Isupp$ are defined as the y and x-intercept that give the best fit between a straight line and the $\Jideal(\ih)$ function. The following values of $\alpha$ and $\beta$ were obtained by simply fitting a line to the $\Jideal(\ih)$ function directly: $\eta=2$: $\alpha=1.25$, $\beta=0.88$ and $\eta=3$: $\alpha=1.12$, $\beta=0.70$.

\newpage
\printbibliography